\documentclass[twocolumn,tighten,times,twocolappendix]{aastex631}

\usepackage{amssymb,amsmath}
\usepackage[inline]{enumitem} 
\usepackage{soul}
\usepackage{natbib}
\usepackage{color}
\usepackage{threeparttable}
\usepackage{longtable}
\usepackage{booktabs}

\usepackage{ulem}

\shorttitle{Huge magnetic toroids}
\shortauthors{J. Xu and J. L. Han}

\begin{document}

\title{The huge magnetic toroids in the Milky Way halo}

\correspondingauthor{hjl@nao.cas.cn}

\author{J. Xu}
\affil{National Astronomical Observatories, Chinese Academy of Sciences,
     A20 Datun Road, Chaoyang District, Beijing 100101, China}
\affil{School of Astronomy and Space Sciences,
       University of Chinese Academy of Sciences, 
       Beijing 100049, China}

\author{J. L. Han}
\affil{National Astronomical Observatories, Chinese Academy of Sciences,
     A20 Datun Road, Chaoyang District, Beijing 100101, China}
\affil{School of Astronomy and Space Sciences,
       University of Chinese Academy of Sciences, 
       Beijing 100049, China}

\begin{abstract}
{The magnetic fields in our Milky Way can be revealed by the distribution of Faraday rotation measures (RMs) of radio sources behind the Galaxy and of radio pulsars inside the Galaxy. Based on the antisymmetry of the Faraday sky in the inner Galaxy to the Galactic coordinates, the magnetic field toroids above and below the Galactic plane with reversed field directions exist in the Galactic halo and have been included in almost all models for the global magnetic structure in the Milky Way. However, the quantitative parameters, such as the field strength, the scale height, and the scale radius of the toroids are hard to determine from observational data. It has long been argued that the RM antisymmetry could be dominated by the local contributions of the interstellar medium. Here we discount the local RM contributions measured by pulsars mostly located within 5 kpc and get the first quantitative estimate of the size of magnetic toroids in the Galactic halo. They are huge, starting from a Galactocentric radius of less than 2~kpc to at least 15~kpc without field direction reversals. Such magnetic toroids in the Galactic halo should naturally constrain the physical processes in galaxies.}
\end{abstract}


\section{Introduction}
\label{sect1}

Magnetic fields are ubiquitous in our Milky Way and play a significant role in the evolution of molecular clouds and star formation \citep{cru12} and the transport of cosmic rays \citep{ps03,ls11}. However, the measurements of the Galactic magnetic fields have been merely limited to some components of the three-dimensional vector fields in limited regions \citep{han17}. It is very challenging to outline the global magnetic structure of the Galaxy from limited available measurements. 
The starlight polarization \citep{hei96,cppt12,ccc+20} is the accumulated result of selective extinction by nonspherical dust grains aligned by interstellar magnetic fields between stars and us, measuring the magnetic field orientations perpendicular to the line of sight. So does the polarized thermal emission from dust grains in molecular clouds \citep{cru12,paa+16b}. The polarized diffuse radio emission \citep{blw+13, paa+16} outlines the orientation of the magnetic fields in the sky plane, which comes from the weighted sum of all Faraday-rotated synchrotron emission all along a line of sight. The Zeeman splitting measures only the in-site field in the dense gas region from the spectral lines \citep{cwh+10}. The Faraday rotation of polarized radio sources measures the line-of-sight magnetic field component weighted by the electron density ($n_e$) along the path between the source and us, through rotation measure (${\rm RM}$):  
\begin{equation}
{\rm RM}=0.812\int_{\rm us}^{\rm Source}n_e \mathbf{B} \cdot {\rm d}\mathbf{l}.
\end{equation}
Here the ${\rm d}\mathbf{l}$ is the element path vector along the line of sight. For pulsars inside our Milky Way or the fast radio bursts in other galaxies, thermal electrons in the path cause more delays for the pulsed emission at lower frequency, so that the integration of electrons can be measured as the dispersion measures (${\rm DM}$) given by 
$ %
{\rm DM}=\int_{\rm us}^{\rm Source} n_e {\rm d}l. 
$ %
At present, the Galactic distribution of electron density in the Milky Way has been modeled from the pulsar DMs and independently measured distances \citep{cl02,ymw17}. Based on the RMs and DMs of a large number of pulsars as well as the RMs of background radio sources at the lower Galactic latitudes, e.g. within 8 degrees from the Galactic plane, the large-scale magnetic fields\footnote{The magnetic fields of the Milky Way have large-scale regular components and small-scale random turbulent components \citep[e.g.][]{han17}. The large-scale fields are generally coherent on scales larger than a few kpc with a strength of some $\mu$G, while the turbulent random fields are much stronger on scales of less than 100 pc. We work in this paper on the global large-scale structure of the magnetic fields in the disk and halo of our Galaxy.} in the Galactic disc have been revealed to follow spiral arms with reversals between the arms and inter-arm regions \citep{hml+06,hmvd18,xhwy22}.

\begin{figure*}[t!]
  \centering 
  \includegraphics[width=0.88\textwidth] {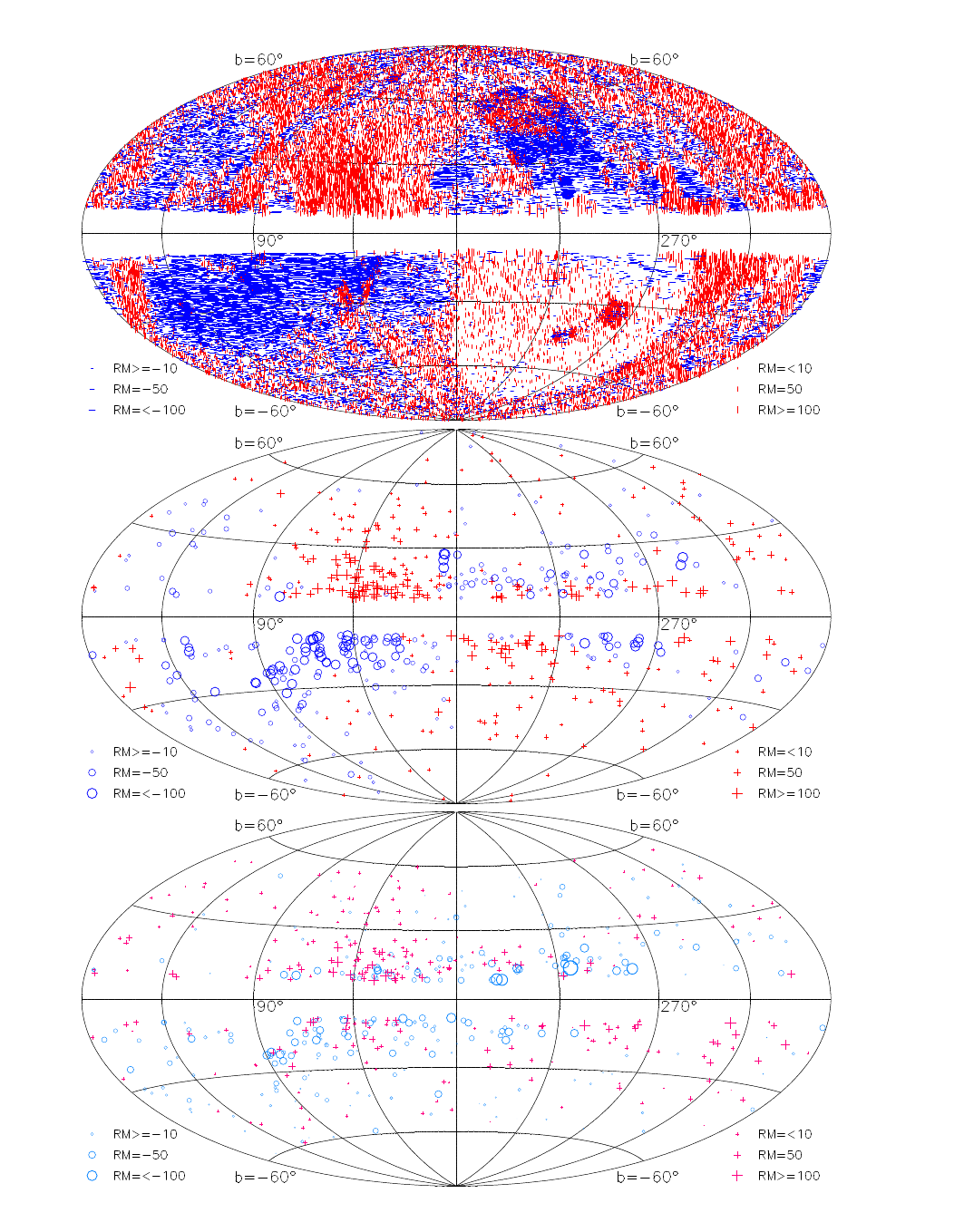} 
  \caption{The sky distribution of Faraday rotations of background radio sources and pulsars. {\it In the top panel}, the RM sky constructed by the RMs of background radio sources at high Galactic latitude of $|b|>8^{\circ}$ from data in the updated RM catalog \citep{xh14}. Any sources with an RM deviating more than 3 times the standard deviations of their 30 neighborhoods have been discarded.  {\it In the middle panel}, the RM distribution of 634 pulsars at high Galactic latitude of $|b|>8^{\circ}$. These pulsars in globular clusters and Magellanic Clouds are excluded.  {\it In the bottom panel}, the local-discounted RMs in the direction of 543 pulsars with an uncertainty smaller than 15 rad m$^{-2}$. The sizes of symbols in all panels are proportional to the square root of the RM magnitudes, with limits of 10 and 100 rad~m$^2$ as marked. 
    }  
\label{RMsky8}
\end{figure*}

The magnetic fields in the Galactic halo are featured by polarized radio emission in the high Galactic latitude sky \citep{blw+13, paa+16} and the averaged RM distribution of radio sources \citep{hmbb97, hmq99, tss09, rmsky2022}. The observed quantities at the high Galactic latitudes, however, are the sum of the contributions from the small-scale local features in the Galactic disk and the tenuous medium in the vast halo space. The local features from the interstellar medium are dominant in some sky regions \citep{hmg11,slg+15}, and they should first be discounted when one tries to derive the global magnetic field structure in the vast Galactic halo.

Based on a great increase of RM data of background radio sources and also of a large number of pulsars in recent years, we investigate in this paper the RM distribution from the Galactic halo beyond pulsars to the Galactic outskirts and quantitatively figure out the properties of the magnetic toroids in the halo. The arrangement of our paper is as follows. We introduce the RM data in Section~\ref{sect2} and analyze the local-discounted RM sky in Section~\ref{sect3}. Modeling for the Galactic halo magnetic field is given in Section~\ref{sect4}. Finally, we present our discussion and conclusion in Section~\ref{sect5}.

\begin{figure}[t]
  \centering \includegraphics[angle=-90,width=0.38\textwidth]{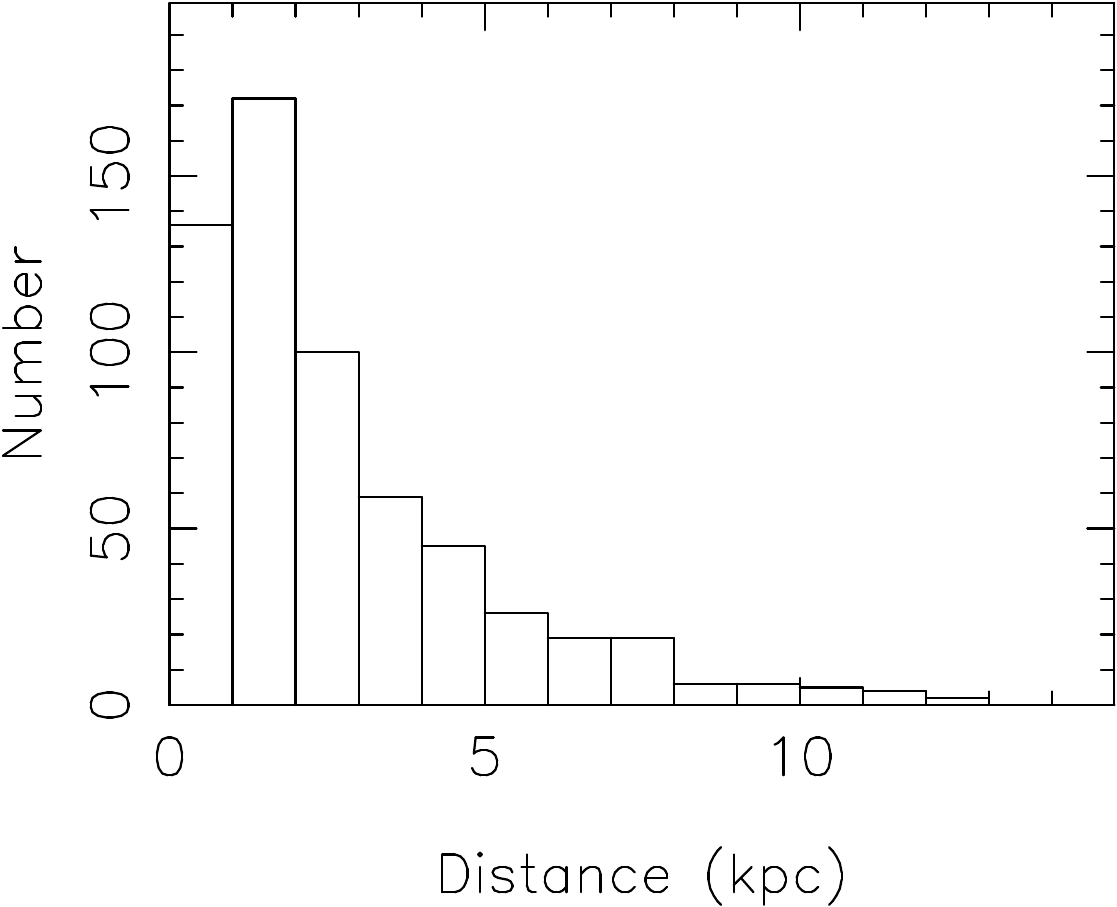}
  \caption{Distance distribution of 634 pulsars at high Galactic latitude of $|b|>8^{\circ}$.
    }  
\label{dist8}
\end{figure}

\begin{figure*}[t]
  \centering \includegraphics[width=0.70\textwidth] {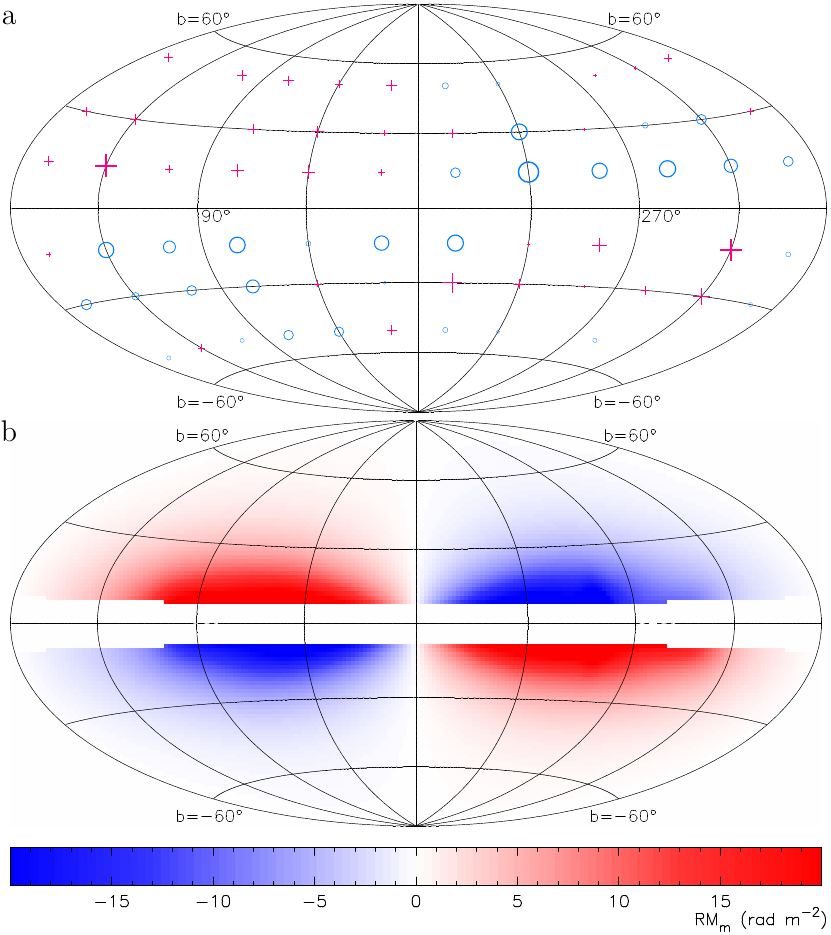}
  \caption{The observed and modeled mean RM sky after the local RM contribution is discounted. {\it In the top panel (a),} the distribution of median values of local-discounted RMs in the 72 binned sky regions in the latitudes of $|b|=(8^{\circ},20^{\circ})$, $(20^{\circ},40^{\circ})$ and $(40^{\circ},60^{\circ})$ and 12 longitude ranges of $30^{\circ}$ each. The discounted RMs with an uncertainty larger than 15 rad m$^{-2}$ and those for pulsars in globular clusters are all discarded. For clarity, the sizes of symbols are enlarged triply compared to those in Figure~\ref{RMsky8}.
  {\it In the bottom panel (b),} the simulated local-discounted RM sky at $|b|>8^{\circ}$ calculated by using the optimized model for the magnetic field toroids in this paper, integrated the Faraday effect from 3~kpc of the Sun to the outskirts of our Milky Way galaxy.
    }  
\label{RMsky8residual}
\end{figure*}

\section{Faraday rotation measure data}
\label{sect2}

We collect the Faraday rotation measure data sample for background radio sources and pulsars. The RMs for background radio sources are taken from the NVSS RM catalog \citep{tss09} and the updated RM catalog compiled from the literature by authors \citep{xh14}. We obtain a sample of RMs of 59233 background radio sources, with the overall mean number density of 1.5~$deg^{-2}$ (see Figure\ref{RMsky8}). For pulsars, except for the RMs collected from radio observations all over the worlds \citep{mhth05}, we also have got our own pulsar RM data for faint pulsars \citep{xhwy22,whx+23} observed by using the Five-hundred-meter Aperture Spherical radio Telescope (FAST) \citep{nan08}. In total, we have RMs of 634 pulsars at the Galactic latitudes greater than 8 degrees (see Figure\ref{RMsky8}).

\subsection{Faraday Rotation Measure Data of Background Radio Sources behind the Milky Way}

The RMs for background radio sources are taken from the NVSS RM catalog \citep{tss09} and the updated RM catalog compiled from the literature by authors \citep{xh14}. 

The NVSS RM catalog has 37,543 sources which are derived with two narrow-band polarization data at 1.4 GHz, almost uniformly covering the sky area above a declination of $-40^{\circ}$. The updated RM compiled catalog of \citet{xh14} now has a total of 24168 sources, including the original 4553 sources compiled in 2014 and new RM data obtained mainly by the ATCA broadband spectropolarimetric survey \citep{agff15}, the Parkes S-band Polarization All Sky Survey \citep{lrf+16}, the low-frequency Polarised GLEAM Survey \citep{rlv+18,rgs+20}, the S-PASS/ATCA of the southern sky \citep{scw+19}, the Canadian Galactic Plane Survey \citep{vbo+21}, the LOFAR Two-metre Sky Survey \citep{osh+23}, the low-band Rapid ASKAP Continuum Survey \citep{tml+23}, the Apertif science verification campaign \citep{aba+22}, some regional targeted observations \citep{kpg+17, bhm+19, sls+19, mmob20, jmh21, lmg+21, lmm+22}, supplemented by some sporadic RM observations.

Finally, we have a sample of RMs of 59233 background radio sources. In the southern sky of $\delta<-40^{\circ}$, there are 3192 RMs mostly obtained by the S-PASS/ATCA project \citep{scw+19}, and hence the number density is lower, only about 0.45 $deg^{-2}$. 

The RM distribution is cleaned by discarding any RMs deviating from their 30 nearby neighbors by more than three standard deviations, following \citet{hmbb97,hmq99}. The cleaned RM sky at high Galactic latitude of $|b|>8^{\circ}$ is shown in the top panel of Figure~\ref{RMsky8}.

\subsection{Faraday Rotation Measure Data of Pulsars}

Currently, 3389 pulsars have been discovered and listed in the updated pulsar catalog \citep{mhth05}\footnote{https://www.atnf.csiro.au/research/pulsar/psrcat/ (version 1.70)}, and 1493 of them have published RMs. Among 1126 pulsars in the high Galactic latitude of $|b|>8^{\circ}$, and 517 of them have measured RMs. The remaining pulsars are too faint to be measured efficiently for RMs by other radio telescopes. We carried out FAST polarization observations in two projects [PT2020\_0164 and PT2021\_0051] from 2020 October to 2021 November, and we determined new RM values for 134 faint halo pulsars \citep{xhwy22}. The 
pulsar data density is improved in the left sky part (i.e. the first and the second quadrants of our Galaxy). We also collect a few RMs of halo pulsars by other FAST projects \citep{whx+23,zhx+23} and new measurements from other telescopes \citep{pkj+23,rvwc23}, and finally obtain a sample of RMs of 634 pulsars. Among them, 409 pulsars  (64.5\%) have distances less than 3 kpc from the Sun, and 512 pulsars (80.8\%) are located within 5~kpc. The distances are estimated by the electron density model YMW16, and the distribution is shown in Figure~\ref{dist8}. 

The RM distribution of 634 pulsars at the Galactic latitudes greater than 8 degrees is shown in the middle panel of Figure~\ref{RMsky8}. The FAST measurements improve the data coverage in the first and the second quarters of our Galaxy. We can see the antisymmetric distribution of the pulsar RMs in the inner Galaxy, i.e. $270^{\circ}<l<90^{\circ}$, which means that the local interstellar medium contributes dominantly to the observed magneto-ionic structure in the sky.

\section{The halo field modeling for the local-discounted RM sky}
\label{sect3}

Here we consider the observed RMs of the background radio sources, which consist of contributions from the medium local to a radio source, from the intergalactic medium between a source and the Milky Way, and from the interstellar medium between the outskirt of the Milky Way to us. The common term for a group of sources nearby in the sky is the RM foreground from our Milky Way, because the source local RMs are unknown and unrelated to each other, and hence probably random; the RMs from the intergalactic medium are probably small \citep{xh14b,cvo+22} and also radially random because of their different distances. 

As shown in the top panel of Figure~\ref{RMsky8}, the RM sky reflects the RM contribution from our Milky Way, which is obtained by filtering out the outstanding RMs of background radio sources deviating from their 30 nearby neighbors by more than three standard deviations \citep[see][]{hmbb97,hmq99}. Because we work on the magnetic fields in the Galactic halo, we exclude the RM data of radio sources and pulsars at the Galactic latitudes of less than 8 degrees, because they are dominated by the interstellar medium in the Galactic disk. The RM distribution in the high Galactic latitude sky is coherent on scales of several degrees to tens of degrees. In the inner Galactic quadrants, i.e. $270^{\circ}<l<90^{\circ}$, the predominant is the antisymmetric distributions of RMs to the Galactic plane and the meridian through the Galactic Centre, although some small anomalous regions stand out for some known local objects. The RM antisymmetry, clearer with more accumulated data in last decades \citep{tss09,xh14,ojr+12,scw+19,hab+22,vgh+23}, has been taken as being evidence for the magnetic toroids in the Galactic halo, i.e. the large-scale azimuthal fields with opposite directions above and below the Galactic plane \citep{hmbb97,hmq99}. In the outer Galaxy, i.e. $270^{\circ}>l>90^{\circ}$, the RM distribution is more symmetric below and above the Galactic plane, dominant by the local dense magneto-ionized medium or distinct objects in the Local arm and the Perseus arm \citep{mmg+12, dwt+22}.

Such magnetic toroids in the Galactic halo have been widely adopted in the models for the global magnetic field structure in the Milky Way \citep{ps03,srwe08,jf12,ft14,svt22}. However, the quantitative parameters of the halo fields such as the field strength and the size of toroids have not yet been constrained by observational data. 

To investigate the properties of magnetic toroids in the Galactic halo, we take the RM measurements of pulsars as being the RM contributions of the local interstellar medium, and then discount them from the RM sky. Despite the perturbations caused by distinct local objects \citep{wfl+10, slg+15}, the pulsar RM distribution in Figure~\ref{RMsky8} at the Galactic latitudes greater than 8 degrees shows the antisymmetry to the Galactic coordinates \citep{hmbb97,hmq99}, which is also naturally produced by the magnetic toroids with reversed directions in the local interstellar medium. Now, the key question is whether or not such toroidal magnetic fields exist in the entire vast Galactic halo beyond these pulsars.

\subsection{Local-discounted RM sky}

\begin{table*}
\centering
\footnotesize
\caption{The local-discounted Galactic RMs in the format of ``Median RM (positive number of RMs /negative number of RMs)'' are listed for 72 sky binned regions (see the upper panel of  Figure~\ref{RMsky8residual}).}
\label{numRM}
\setlength{\tabcolsep}{5pt}
\renewcommand{\arraystretch}{1.3}
\begin{tabular}{c|cccccc} 
\hline
  Galactic   &\multicolumn{6}{c}{Galactic latitude bins}\\
  longitude bins  & ($40^{\circ},60^{\circ}$) & ($20^{\circ},40^{\circ}$) & ($8^{\circ},20^{\circ}$) &  ($-20^{\circ},-8^{\circ}$) & ($-40^{\circ},-20^{\circ}$) & ($-60^{\circ},-40^{\circ}$) \\
\hline
($150^{\circ}<l<180^{\circ}$)& 6.0(3/0) & 6.2(5/0) &7.3(5/2)   &1.7(4/3)   &-9.4(1/7)  &-1.5(0/2) \\
($120^{\circ}<l<150^{\circ}$)& -- (1/0) & 7.7(5/2) &38.0(2/1)  &-20.0(1/6) &-4.8(1/4)  &4.4(4/3)  \\
($90^{\circ}<l<120^{\circ}$) & 7.8(2/1)& -- (1/0) &4.8(2/1)   &-12.3(1/4) &-8.2(2/4)  &-1.5(3/5) \\
($60^{\circ}<l<90^{\circ}$)  & 7.8(5/2) & 7.0(6/2) &13.2(8/4)  &-20.9(4/13)&-14.5(6/13)&-7.8(1/3) \\
($30^{\circ}<l<60^{\circ}$)  & 5.0(5/0) & 11.0(19/4)&13.5(22/12)&-2.1(12/13)&3.8(6/4)   &-7.6(0/4) \\
($0^{\circ}<l<30^{\circ}$)   & 9.2(3/1) & 3.0(7/4) &3.3(22/16) &-17.7(3/14)&-0.9(3/5)  &7.6(3/2)  \\
\hline
($330^{\circ}<l<360^{\circ}$)& -3.3(0/2)& 6.1(5/1) &-8.1(6/17) &-21.9(2/11)&29.2(4/0)  &-2.3(1/4) \\
($300^{\circ}<l<330^{\circ}$)& -1.4(1/4)&-21.5(0/6)&-34.7(4/10)&1.1(6/3)   &7.1(3/1)   &-1.0(0/2)  \\
($270^{\circ}<l<300^{\circ}$)& ---(0/1) &0.8(4/3)&-19.7(2/8) &15.4(11/0) &0.5(5/5)  &---(1/0)  \\
($240^{\circ}<l<270^{\circ}$)& 0.8(3/2)&-2.7(3/3) &-22.4(0/2) &---(0/0)   &5.2(3/1)   &-1.5(1/2)  \\
($210^{\circ}<l<240^{\circ}$)& 1.1(2/2) &-8.5(2/7) &-15.1(0/2) &36.1(7/1)  &21.9(6/0)  &---(0/1)  \\
($180^{\circ}<l<210^{\circ}$)& 4.3(3/2) &4.0(4/1)  &-8.3(1/2)  &-2.1(2/2)   &-1.4(1/2)  &---(0/0)  \\
\hline
\end{tabular}
\end{table*}

We obtain the local-discounted RM sky, as shown in the bottom panel of Figure~\ref{RMsky8} by subtracting the pulsar RMs from the medians of RMs of the nearest 30 background radio sources around the directions of pulsars. The median-finding procedure can naturally avoid the disorders caused by outliers of the RMs of background radio sources in the averaging processing, hence giving the best values for the Galactic RMs. In the south sky of the declination $\delta<-40^{\circ}$ the number density of background sources with known RMs is relatively small, so 30 sources are found from a relatively larger sky region, but a limit is set for the largest smooth radius of $5^{\circ}$. The data for 91 pulsars have been excluded because the local-discounted RM values have a large uncertainty\footnote{The uncertainty of local-discounted RM is calculated from square root of quadrature sum of the uncertainty of pulsar RMs and the standard error of the median RM of background radio sources. In addition to the measurement uncertainty of pulsar RMs, the RM of a pulsar contains the RM fluctuation due to the randomness of the local interstellar medium. Such probable RM fluctuations have naturally adhered to the local-discounted RMs, which will be smoothed largely during the statistics for many small sky regions.} of more than 15 rad m$^{-2}$ or those for pulsars are in globular clusters. Such local-discounted RM values in the bottom panel of Figure~\ref{RMsky8} represent the RM contribution from the interstellar medium in the Galactic halo beyond pulsars, i.e. the medium between pulsars and the Galactic outskirts. The local RM contributions from nearby objects such as a local HI bubble \citep{wfl+10} or the Region A \citep{sk80} are then largely discounted because pulsars are generally located much behind these bubbles, more distant than 200 pc.

To analyze the weak toroidal magnetic fields in the Galactic halo from the local-discounted RM sky described by the local-discounted RMs with some certain uncertainty in a limited number of lines of sight, we divide the sky region of $8^{\circ}<|b|<60^{\circ}$ into several regional ``bins'' and then make statistics: 3 bins in the Galactic latitudes of $|b|=(8^{\circ},20^{\circ})$, $(20^{\circ},40^{\circ})$ and $(40^{\circ},60^{\circ})$ and 12 bins along the Galactic longitude with $30^{\circ}$ each from $0^{\circ}$ to $360^{\circ}$. And then we make the statistics, simply counting the number of positive/negative values and calculating the median in each bin. The uncertainties of the local-discounted RMs are then largely suppressed. In the upper panel of Figure~\ref{RMsky8residual} and Table~\ref{numRM}, one can then see that the antisymmetric distribution of RM signs is not limited in the inner Galaxy but also extended to the anti-center of the outer Galaxy, though the amplitude of local-discounted RMs is relatively small, a few rad m$^{-2}$ in general, and more data are desired for better statistics in future. After the RM contribution from the local objects and in the Local arm and the Perseus arm has been discounted, the antisymmetric pattern {\it in the outer Galaxy} is also remarkably standing out, and the RM values around the Galactic latitude of $\pm30$ degree in both hemispheres should follow a sinusoidal wave along the longitude \citep[cf. ][]{dwt+22}. This is strong evidence for the existence of the toroidal magnetic fields in the vast Galactic halo farther than in the local limited by pulsars, which should extend to a much larger Galactocentric radius than the solar orbit. 

\subsection{Models for the magnetic toroids in the Galactic halo}
\label{sect4}

The striking antisymmetric distribution of RMs in the inner Galaxy has been interpreted as an indication of toroidal magnetic fields in the Galactic halo with opposite field directions above and below the Galactic plane \citep{hmbb97,hmq99}. The antisymmetry of local-discounted Galactic RMs shown in the bottom panel of Figure~\ref{RMsky8} is strong evidence of toroidal magnetic fields in the vast Galactic halo. The pattern keeping antisymmetric in the outer Galaxy implies that toroidal magnetic fields exist from a small to large Galactocentric radius, crossing over the solar orbit. We explore the properties of the magnetic toroids and then make models for the toroids to reproduce the RM distribution. We adopt the Galactic electron density model given by YMW16 \citep{ymw17}, in which the outskirt of the Milky Way is set to be at $R_G=20$~kpc and $z_G=10$, 
and the Galactocentric distance of the Sun is taken as an updated value of 8.3 kpc \citep{brm+11}. The model includes a thick disk, a thin disk, and other components. The thick disk has a scale height of 1.673~kpc, comparable to the thick disk component decomposed from the Galactic synchrotron sky \citep{bkb85}. Noticed that the other popular density model is the NE2001 model \citep{cl02}, which also includes a thick disk with a similar electron density at heights of $z>2$~kpc to the YMW16 model. To mimic the effect of magnetic fields in the thick disk or the halo and to reproduce the local-discounted RM distribution, we ignore all magnetic fields in the Galactic disk. For the magnetic fields in the toroids, the field directions should reverse magnetic field direction below and above the Galactic plane, so we adopt $B_{\phi}(R,\phi,z) = sign(z)B(R,z)$. Here, $sign(z)$=1 for $z>0$ and $-1$ for $z<0$, $B(R,z)$ is the field strength as a function of the Galactocentric radius $R$ from the Galactic center and the vertical height $z$ from the Galactic plane.

We wish to constrain parameters of the magnetic toroids, such as the strength of the magnetic fields and also the scale radius $R_0$ and scale height $Z_0$. Some hints are recently available from the new measurements by \citet{xh19} and \citet{xhwy22}: The DM and RM measurements of high latitude pulsars suggest the local magnetic field to have a strength of around 2$\mu$G. The scale height of the magnetic fields is $z_0$ at least 2.7 kpc if the fields decline exponentially in the vertical direction. 

In the following, we make models to reproduce the antisymmetric local-discounted RM distribution. To qualify the possible magnetic field reversals in the magnetic toroids and to demonstrate the affected sky region of the RM distribution, we first take the simplest case and see the results. Then we construct a reasonable model to fit the antisymmetry of local-discounted Galactic RMs. Noting that the modeling here only characterizes the large-scale component of the Galactic magnetic fields, not the random turbulent magnetic fields which exist in small scales and therefore are beyond the scope of this paper.

\begin{figure*}
  \centering \includegraphics[width=0.9\textwidth,trim=70 0 50 0,clip] {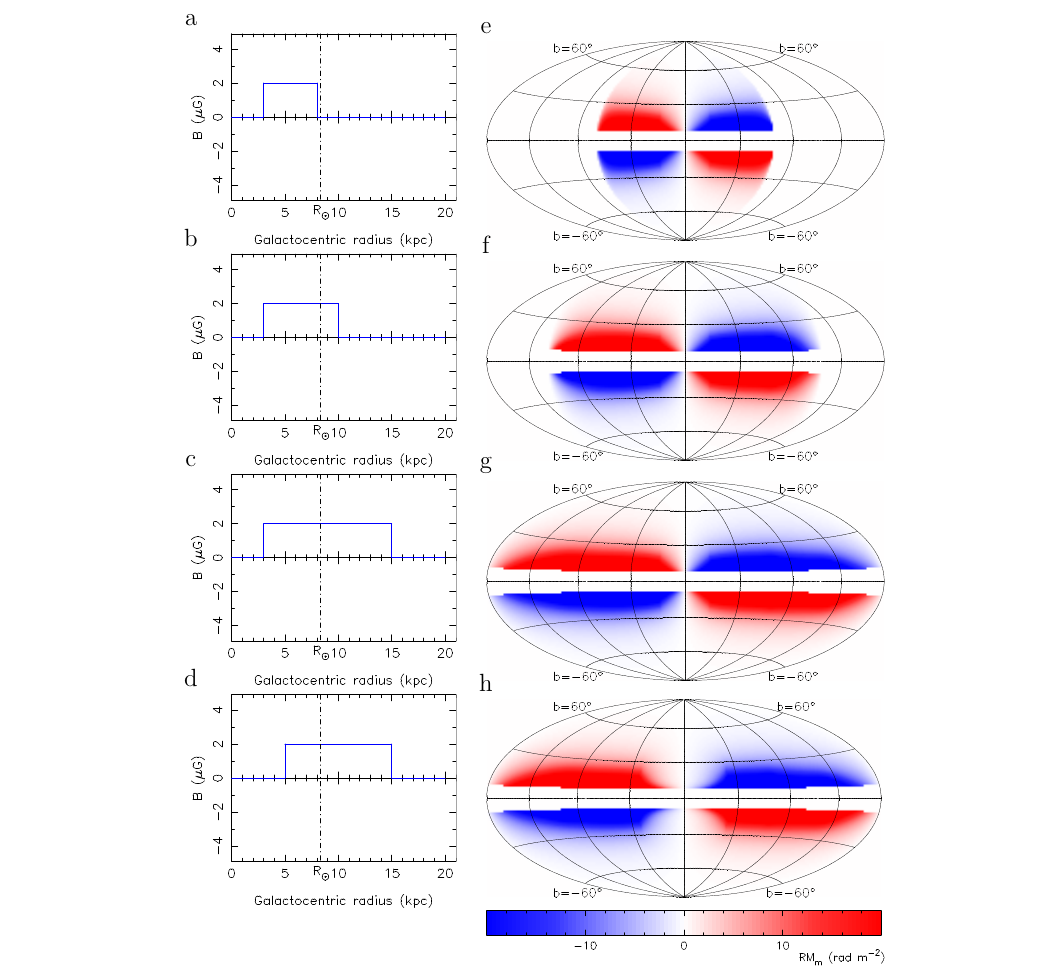} 
\caption{The simulated RM distribution at high Galactic latitudes by qualitative field models for a constant field strength. The Galactic electron density model, YMW16 \citep{ymw17}, has been used for calculation. The constant magnetic fields are assumed for the magnetic toroids with cutoffs of ($R_1$, $R_2$) = (3.0, 8.0), (3.0, 10.0), (3.0, 15.0), (5.0, 15.0) kpc as shown in the panels of (a) - (d), and the RM antisymmetric RM distributions from these models are given in the panels of ((e) -- (h). The modeled RMs within 3~kpc from the Sun are not accounted for in these models.}  
\label{simu11}
\end{figure*}

\begin{figure*} 
  \centering \includegraphics[width=0.9\textwidth,trim=70 0 50 0,clip] {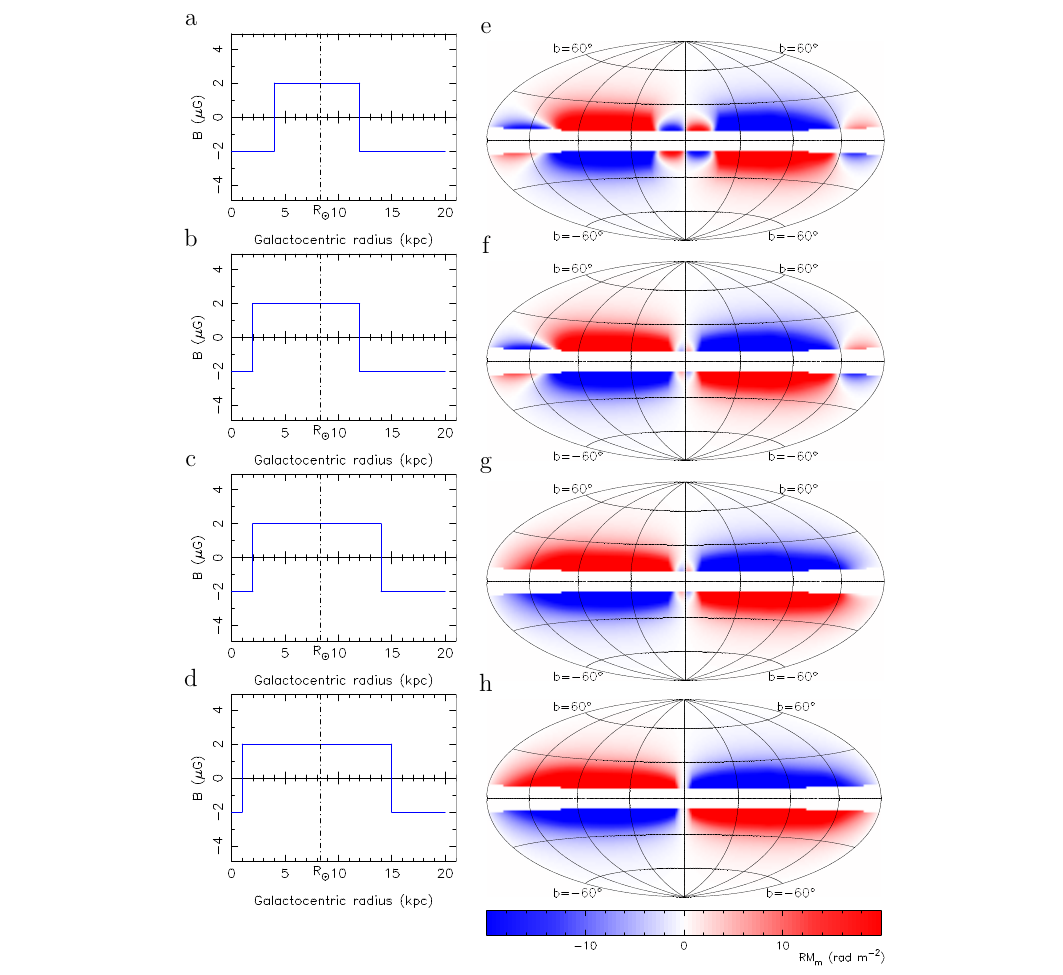} 
\caption{The same as Figure~\ref{simu11} but for a constant halo field strength with direction reversals in some radius ranges}. 
\label{simu22}
\end{figure*}

\subsubsection{Qualitative models with a constant field strength without radial reversals}

The simplest case is a constant field strength without reversals along the Galacto-radius inside the toroids with a given scale-height. The field strength varies with $z$ by  
$ 
B(R,z) = B_{0}exp(-|z|/z_0), 
$ 
here, we take $z_0=2.7$~kpc and $B_0 = 2$~$\mu$G. The toroids have the inner radius cutting off at $R_1$ and the outer radius cutoff $R_2$. As mentioned above, the outskirt boundary is set to $R_G = 20$~kpc and $z_G = 10$~kpc. The RMs produced by the local interstellar medium within 3~kpc from the Sun are not counted. 

The results for some different settings of $R_1$ and $R_2$ are shown  in Figure~\ref{simu11}. The magnetic toriods for these different Galacto-radius ranges give the different boundary for the antisymmetry. The outer radius cutoff determines the outer boundary of the antisymmetry, and the inner radius cutoff set the sharpness of the RM transition from the meridian of the Galactic center. To get the antisymmetry extended to the anticenter of the Milky Way, the outer radius cutoff must go to $R_2 \sim 15$ kpc. Similarly, a larger inner cutoff radius causes a larger area for the RM null around the direction of the Galactic center. 

From such qualitative models,  one can conclude that magnetic toriods must extend to the outer Galaxy, at least to $R_2$=15 kpc. The inner cutoff radius should be smaller than $R_1$=3 kpc, if there is no field reversals in the halo.  

\subsubsection{Qualitative models for a constant field strength but with radial reversals}

Now we consider possible field reversals of directions in the  magnetic toroids in some Galacto-radius ranges. Between the radius range of $R_1$ and $R_2$, the field strength is 
$
B(R,z) = B_{0}exp(-|z|/z_0),
$ 
and outside this range, 
$B(R,z) = -B_{0}exp(-|z|/z_0) $. 
The results for the distribution of the local-discounted RMs are  shown in Figure~\ref{simu22}. Corresponding RM sign changes near the Galactic center or the anticenter appear if $R_1 > 1$~kpc or $R_2 < 15$~kpc. 

We then conclude there should not be any large-scale field direction reversals in the magnetic toroids between the Galacto-radii 1 kpc and 15 kpc.

\subsection{The best model for the magnetic toroids in the Galactic halo}

The qualitative models above reveal the interplay between the large-scale field reversals and the regions for sign-changes of RM distributions, in spite of any field strength.

To get the best model for the Galactic halo magnetic field is not an easy task. It is difficult to completely separate the RM contributions from the Galactic halo and from the Galactic disc, because the Sun is located in the mid-plane of the disc and because the lines of sight to radio sources in the halo regions certainly traverse some turbulent disc regions. The RM contribution from the Galactic disc can cause  clear fingerprints in northern and southern hemisphere \citep{dwt+22}. The observed RMs in the outer Galaxy are mainly contributed by nearer Galactic disc regions. Nevertheless, the local-discounted RMs greatly diminish contribution from the disc magnetic field component in front of pulsars. 

Here we develop a halo model to fit the basic features in the observed local-discounted Galactic RM distribution beyond pulsars, which contains the random contributions from the turburent fields. We consider more practical parameters for characterizing the large-scale magnetic toriods. In order to set the halo field strength to be zero at $z =0$, we add a factor of $\frac{|z|}{z_{0}}$ to the exponential vertical dependence defined by the qualitative models. For the radial field strength profile of the magnetic toroids, we find that the Gaussian radial profile can match the data better than an exponential profile or other formats. Therefore we take 
a Gaussian function $\exp{\left [ -\left ( \frac{R-R_{0}}{R_{T}} \right )^{2} \right ]}$ to characterize the radial profile of magnetic toroids, with the maximum field strength at $R_{0}$ and a scale radius of $R_{T}$. As seen in Figure~\ref{RMsky8residual}, the largest local-discounted RM values are located at Galactic longitude $l \sim 50^{\circ}-90^{\circ}$, which constrains the Galactocentric radius of the Gaussian profile peak.
Combining these considerations, we have 
\begin{equation}
\begin{split}
B_{\phi}(R,\phi,z) = 
& sign(z)B_{0}\frac{|z|}{z_{0}} \exp{\left (-\frac{|z|-z_{0}}{z_{0}} \right)} \\
& \times \exp{\left [ -\left ( \frac{R-R_{0}}{R_{T}} \right )^{2} \right ]}
\end{split}
\label{gaupro}
\end{equation}
where $B_{0}$ is the field strength, $z_{0}$ is the vertical scale height. As a matter of facts, we also tried other $z$ dependences such as a Lorentzian function $f(z) =\left [ 1+\left (\frac{|z|-z_{0}}{z_{1}} \right )^{2} \right ]^{-1}$ which has more parameters \citep{ps03,srwe08}, and found that the $z$-dependence is not important. We therefore prefer to take a model with less parameters.

\begin{figure}[t]
     \centering \includegraphics[width=0.47\textwidth]{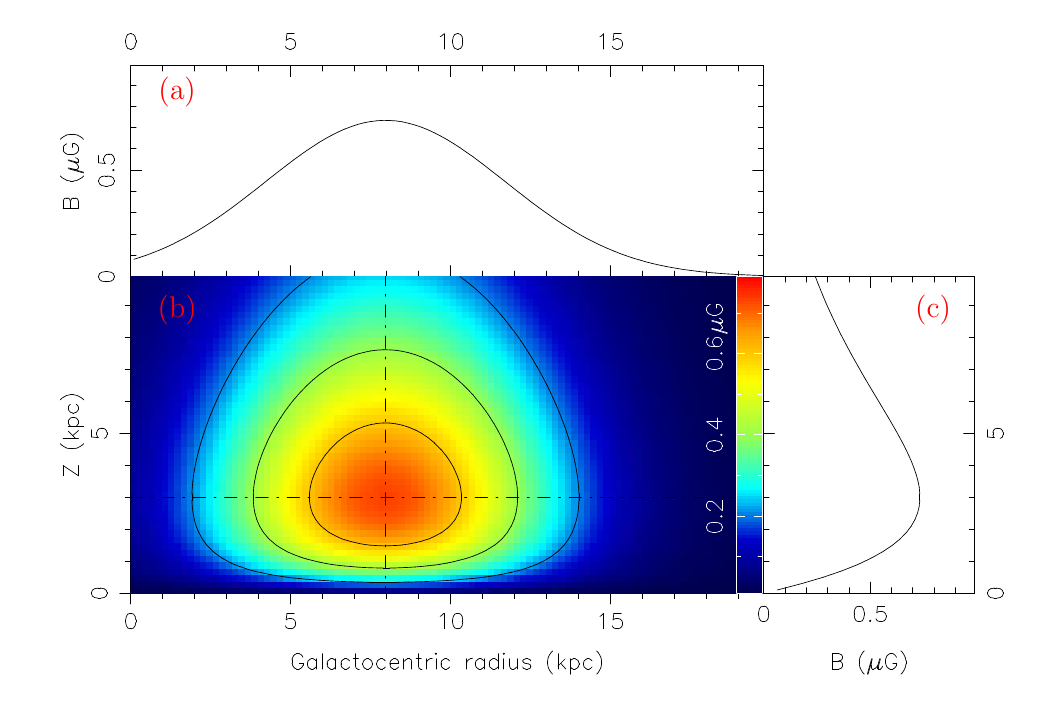}
  \caption{The best model for the magnetic toroids in the Galactic halo. The subpanel (a) shows the field strength versus the Galacto-radius $R$ along the horizontal dot-dashed line in the 2D distribution of the field strength in the $R-z$ plane in the subpanel (b). The subpanel (c) gives the strength versus the vertical distance from the Galactic plane $z$ along the vertical dot-dashed line in the $R-z$ plane. Contours in the subpanel (b) are at levels of 0.2, 0.4 and 0.6~$\mu$G.} 
\label{modelBRz}
\end{figure}

\begin{figure*}
     \centering \includegraphics[angle=-90,width=0.75\textwidth]{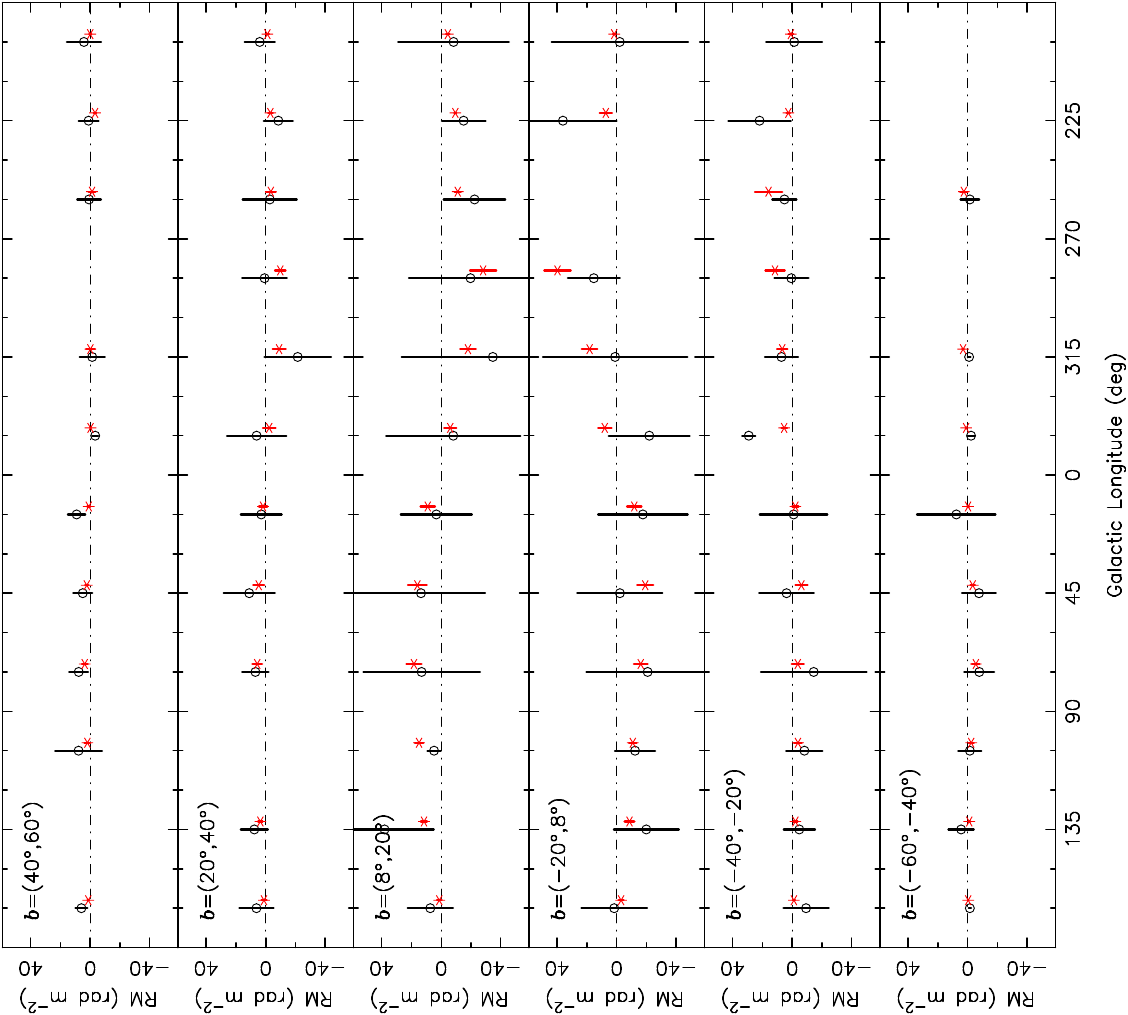}
  \caption{Distribution of the median RMs of local-discounted Galactic RMs beyond pulsars with the Galactic longitudes at several ranges of the Galactic latitude. The median local-discounted Galactic RMs for binned regions along the Galactic longitude are indicated as black circles, with error bars representing the standard deviation from the median. The red asterisks are the median values of RMs calculated from the best-fitted model.} 
\label{modelcompare}
\end{figure*}

\begin{table}
\centering
\caption{Parameters of the optimized model for the magnetic field toroids.}
\label{modelpara}
\begin{tabular}{cccc} 
\hline
$B_{0}$($\mu$G) & $z_{0}$(kpc) & $R_{0}$(kpc) &  $R_{T}$(kpc) \\
$0.73_{-0.12}^{+0.12}$ &  $3.00_{-0.65}$ & $7.97_{-0.93}^{+1.49}$ & $5.31_{-1.75}^{+3.91}$ \\
\hline
\end{tabular}
\end{table}

\begin{figure*}
     \centering \includegraphics[width=0.75\textwidth]{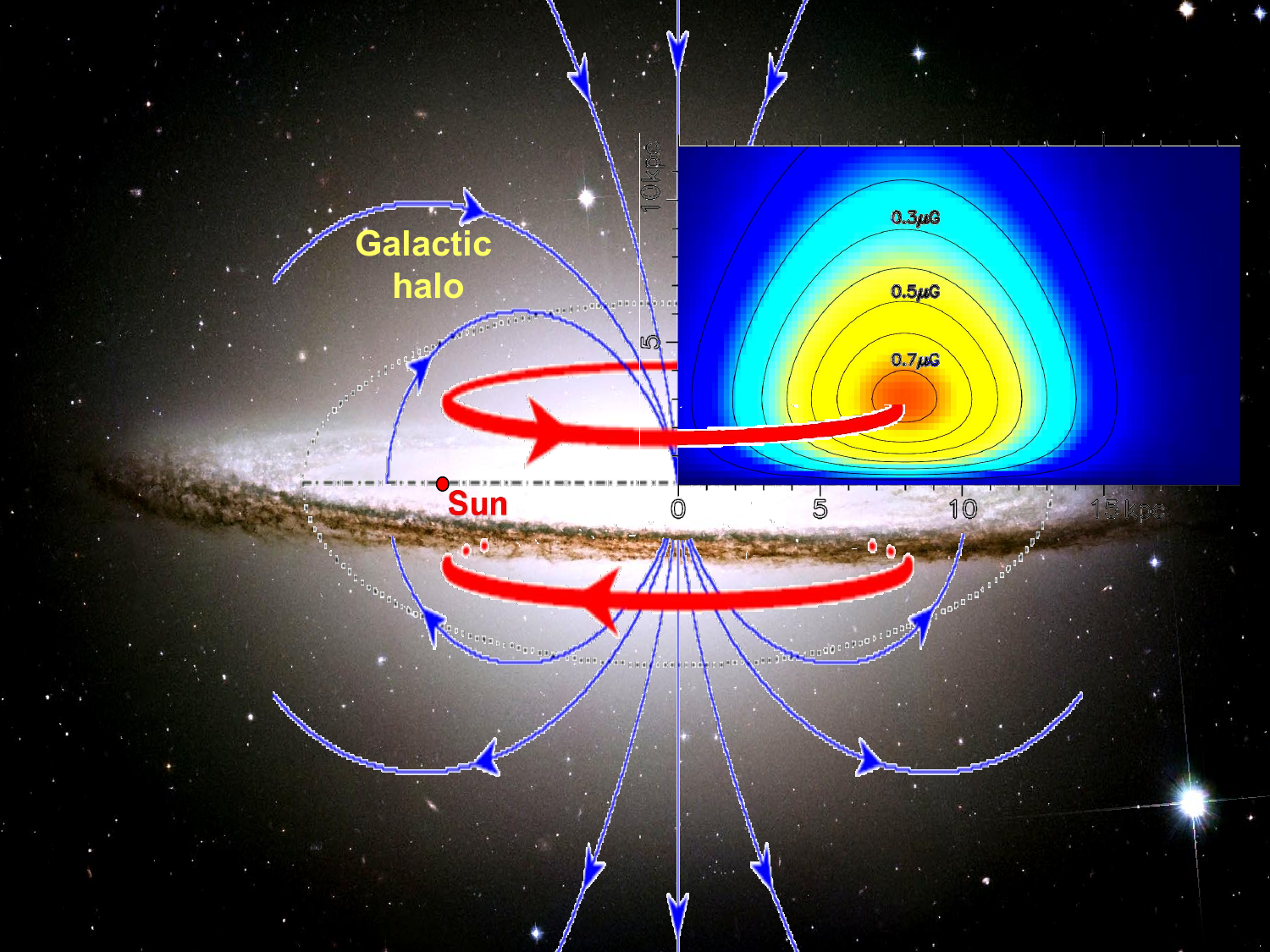}
  \caption{An illustration of the magnetic field structure in the Galactic halo. The background figure is adopted from \citet{han17}, and the best model of the magnetic toroids is obtained from the data fitting in this work. The toroidal fields (red lines) together with the dipolar fields (blue lines) can be naturally created by either differential rotation of vertical magnetic field lines in the concentrated halo \citep{sf87} or the A0 dynamo actions \citep[e.g.][]{wk93}.}
\label{haloB}
\end{figure*}

We performed a model fitting for the magnetic field toroids to the local-discounted Galactic RMs. We take $\chi^2=\sum\limits_{i}^{N}{\frac{\Delta RM_{i}^{2}}{\sigma_{i}^{2}}}$ as an estimator for the fitting quality. Here $N$ is the number of pulsars, $\Delta RM_{i}=(RM_{obs}-RM_{m})_{i}$ is the difference between the model value and observed local-discounted Galactic RM for the $i$th pulsar. The variance $\sigma_{i}^{2}$ is of importance for $\chi^2$ evaluation, which counts for the variance from small-scale random magnetic fields, fluctuations of electron density and other unknown turbulence. 
The local-discounted Galactic RMs has been derived for the RM behind every pulsar with different distances and directions, which include the uncertainty of pulsar RMs and also the RM variance due to the turbulent interstellar medium in front of and also behind the pulsar that is distance-dependent. It is impossible to have many pulsars in similar directions at similar distances for this work, therefore we have to leave all these subtle variance to the subtracted RM values. At present, the $\sigma_{i}^{2}$ from the RM variance of the nearest 30 cleaned background radio sources in the direction of a pulsar is taken as the weight. The {\sc MINUIT} package \citep{jr75} with the {\sc MIGRAD} search routine is adopted to optimize the parameters of the models. 

Noticed that the DM-estimated distances from the Galactic electron density model \citep{ymw17} have a larger uncertainty for more distant pulsars. For 20\% of 634 pulsars located at distance larger than 5 kpc, we set these pulsars at a distance of 5 kpc in the model fit. 

Noticed that in the model there is a degeneracy between the field strength $B_{0}$ and the vertical scale height $z_{0}$, we then set an upper limit $z_{0} \leq $ 3.0 kpc during the fitting process. The final parameters for the best model of the magnetic toroids are obtained and given in Table~\ref{modelpara}, which can produce the minimum $\chi^2=482$ for 543 sightlines with 4 model parameters, or the reduced chi-square $\chi^2_{red}=\chi^2/dof=0.89$. The radial and $z$ dependence of the field strength in the magnetic toroids is shown in Figure~\ref{modelBRz}. The uncertainties for each parameters are derived by using the {\sc MINOS} algorithm in the {\sc MINUIT} package, which finds the positive and negative errors for each parameter where $\chi_{UP}^2=\chi_{min}^2+UP$, while marginalizing over the other $N-1$ paramters. Here $\chi_{min}^2$ is the minimum with respect to all $N$ parameters. The confidence intervals for 68.3\% probability (1 $\sigma$) are obtained for $UP=1$. 

The local-discounted RM sky calculated from the model of magnetic field toroids above and the Galactic electron density model \citep{ymw17} is shown in the lower panel of Figure~\ref{RMsky8residual}. The great antisymmetry over all the sky region, from the meridian of the Galactic center to the anticenter, is reproduced well. A bin-by-bin comparison between the local-discounted Galactic RM and the best-fit model is shown in Figure~\ref{modelcompare}. The figure shows the median values of the observed local-discounted Galactic RM along with the median values of the modeled Galactic RM computed for each source using the best model. The RM variations along Galactic longitude at each latitude range can be well reproduced.

\section{Discussion and Conclusion}
\label{sect5}

The local-discounted RM sky shows a striking antisymmetry to Galactic coordinates in the whole longitude range as seen in Figure~\ref{RMsky8} and Figure~\ref{RMsky8residual}, indicating a huge size of magnetic fields of odd type in the halo.

From the distribution of the local-discounted RMs, we get the basic parameters of toroidal magnetic fields in the Galactic halo by model-fitting. The small amplitudes of the local-discounted RMs result from the combination of weak toroidal magnetic fields at a high latitude and a very low electron density in the vast tenuous halo. As shown in Section~\ref{sect4}, we have tested several possible configurations of the toroidal fields, combined them with the model of the Galactic electron density distribution YMW16 \citep{ymw17}, and found that the magnetic toroids must extend from the Galactocentric radius of less than 2~kpc to more than 15~kpc without any field direction reversals along the radius, otherwise the antisymmetry of RM distribution in the sky can not be reconstructed. 

It is the first time to get the huge size for such dual magnetic toroids in the Galactic halo. The best model with a Gaussian radial profile along the radius and the vertical distance from the  Galactic plane is illustrated in Figure~\ref{haloB}. 

Such magnetic toroids can be generated from the differential rotation of vertical magnetic field lines in the concentrated halo \citep{sf87}. Alternatively, as discussed extensively \citep{wk93, hmbb97, wid02, ft14,srb+19,ntd+20}, they can be generated by the A0 dynamo actions. The A0 dynamo also creates poloidal fields of dipole structure perpendicular to the plane, as seen in Figure~\ref{haloB}. Both the radio filaments near the Galactic Center \citep[e.g.][]{ymc84,yhc04} and the vertical magnetic field in the Solar vicinity \citep{hq94} reflect the existence of poloidal fields.
The dynamo in such a huge size of volume might take time in the order of a Hubble time \citep{bdm+92,srb+19}. Usually, the solutions of the mean-field dynamo are of purely even or odd parity depending on the coupling of strength between the disk and halo \citep{ss90,bdm+92,ms08,msbk10,pm16}. Some numerical simulations \citep{ntd+20} have reproduced a pattern of mixed parity. Therefore the halo magnetic fields may be still in the state of development. A large-scale magnetic helicity \citep{west2020} may play an essential role in the field configuration.  

In the future, more measurements are desired to improve the results further. More RMs of extragalactic sources from large surveys,  such as the Polarization Sky Survey of the Universe's Magnetism (POSSUM) \citep{glt+10}, will improve the coverage and hence source number density of the RM sky, so it can help to improve the accuracy of the Galactic RMs. Finding more pulsars in the Galactic halo and measuring their RMs can better figure out the local RM contributions, which is necessary to separate the halo and local contributions. This can be achieved in future by the Square Kilometer Array (SKA) or its pathfinders \citep[See][]{hvl+15}, for example the Southern-sky MWA Rapid Two-metre (SMART) pulsar survey \citep{bsm+23}, the MeerTRAP survey for pulsars and fast transients with MeerKAT \citep{scd+18} and the LOFAR Tied-Array All-Sky Survey (LOTAAS) for radio pulsars and fast transients \citep{scb+19}, and also our FAST pulsar survey \citep{hww+21}. A refined model for the Galactic distribution of electron density is also a key element for figuring out the nature of the magnetic field toroids.

\begin{acknowledgments}
We sincerely thank the referee for his insightful comments.
We appreciate the discussions and comments from members of the research group on Compact Objects and Diffuse Medium at the National Astronomical Observatories of China. This work is supported by the National SKA Program of China (Grant No. 2022SKA0120103), and the National Natural Science Foundation of China (Grant Nos. 11833009, 11988101, and U2031115).
\end{acknowledgments}

\appendix
\section{ Code for the global Galactic magnetic field model}
  \label{appen} 

We have made good progress on the measurements of the magnetic fields in the Galactic disk and halo. For more convenience in the usage of the best models of the large-scale magnetic field, we assemble the disc field model obtained by \citet{hmvd18} and the halo toroidal field in this paper, so that readers can download the program and work for their own purpose.

The disk field model in \citet{hmvd18} is developed by using low-latitude ($|b|<8^{\circ}$)  RMs of pulsars and extragalactic radio sources behind the disk, which can well reflect the field reversals between the arm and interarm regions. The disk model \citep{hmvd18} assumes logarithmic spiral fields with a radial and z dependence given by 
\begin{equation}
B(R,z)=B_0exp(-R_G/A)exp(-|z|/H).
\end{equation}
Here $R_G$ is the Galactocentric distance. The disk radial scale $A$ and the disk scale height $H$, were taken to be 5.0~kpc and 0.4~kpc, respectively. The spiral fields were assumed to have a pitch angle $\psi=11^{\circ}$. The field strength $B_0=B_s(i)$ for $R_s(i)<R_0<R_s(i+1)$ and 
\begin{equation}
R_0=R_G exp(-\psi \tan \theta).
\end{equation}
Here $\theta$ is the azimuth angle measured counterclockwise from the $+y$ axis. The parameters for $R_s(i)$ and $B_s(i)$ are given in Table 5 of \citet{hmvd18} and also listed in Table~\ref{discmodel}. Here the Galactocentric radius of reversal in the local disc $R_s(6)$ = 8.5 kpc is replaced as 8.16 kpc as done in \citet{xh19}.

\setcounter{table}{0}
\renewcommand{\thetable}{A\arabic{table}}
\begin{table}
\centering
\footnotesize
\caption{Parameters of the disk field model \citep{hmvd18}}
\label{discmodel}
\setlength{\tabcolsep}{5pt}
\renewcommand{\arraystretch}{1.5}
\begin{tabular}{cccccccc} 
\hline
Index $i$ & 1 & 2 & 3 & 4 & 5 & 6 & 7 \\
\hline
$R_s(i)$(kpc) & 3.0 & 4.1 & 4.9 & 6.1 & 7.5 & 8.16 & 10.5 \\
$B_s(i)$($\mu$G) & 4.5 & --3.0 & 6.3 & --4.7 & 3.3 & --8.7 & ... \\
\hline
\end{tabular}
\end{table}

The halo field model in this paper is built based on the local-subtracted RM distribution, describing the toroidal magnetic fields with opposite directions above and below the Galactic plane formulated 
in Equation~(\ref{gaupro}). 
The parameters for the halo model are listed in Table~\ref{modelpara}.

We provide the C code for calculating the magnetic fields at any point inside our Milky Way by using these models. Inside the package, the Cartesian coordinates ($x$, $y$, $z$) originating from the Galactic center and the Galactocentric cylindrical coordinates ($R$, $\phi$, $z$) are used. The $x$ - $y$ plane coincides to the Galactic plane with $x$ paralleling to $l=90^{\circ}$ and $y$ to $l=180^{\circ}$. The Sun is located at ($x$, $y$, $z$)=(0, 8.3, 0). $R=\sqrt{x^2+y^2}$ is the Galactocentric radius, $\phi$ is the azimuth angle starting from $x$-axis and increasing in the counterclockwise direction, $z$ is the distance to the Galactic plane. A user can input the Galactic longitude and latitude and also the distance (in kpc) from the Sun, the package returns three magnetic field components: $B_{r}$, $B_{\phi}$ and $B_{z}$ in the Galactocentric cylindrical coordinates for the disk magnetic fields, the halo magnetic fields and the total magnetic fields, respectively. The package is written in C and is publicly available at http://zmtt.bao.ac.cn/GMF/.

\bibliographystyle{aasjournal}
\bibliography{ref}

\end{document}